\documentstyle[12pt]{article}
\begin{document}
\pagestyle{headings}
\noindent
\large
\begin{center}
Mechanics of a Particle in a Gauge Field\\[8ex]
by\\[2ex]
S. R. Vatsya\\[6ex]
Centre for Research in Earth and Space Science\\[1ex]
York University\\[1ex]
North York, Ontario, Canada~~~~~M3J 1P3\\[5ex]
\begin{tabbing}
Mailing Address: \= c/o ~~Prof. H. O. Pritchard   \\[1ex]
\>Centre for Research in Earth and Space Science  \\[1ex]
\>York University                                 \\[1ex]
\>4700 Keele St., Downsview,                      \\[1ex]
\>Ontario, Canada~~~~~M3J 1P3                     \\[2ex]
\>Phone: \=(416) 736-5363                           \\[1ex]
\>Fax:   \>(416) 736-5936                           \\[1ex]
\>e-mail: huw@gkcl.ists.ca                        \\[7ex]
\end{tabbing}
PACS Categories: 03.65
\end{center}
\newpage\noindent
\begin{center}
Abstract\\[2.5ex]
\end{center}
\normalsize

Classical motion of a charged particle in an electro-magnetic field is
described by the Lorentz equation in terms of the field components. The
field is defined by the infinitesimal gauge group elements associated
with closed curves. The Lorentz equation may be derived from the action
principle with an appropriate Lagrangian. The Lagrangian may also be
used to associate group elements with curves in the space-time manifold.
The action principle is shown here to be an equivalence relation between
the infinitesimal elements so defined for a collection of closed curves
and the identity element. This suggests a natural extension to require
the equivalence of global elements with
the identity and by considering all curves. The
resulting equation, in addition to providing an extension of the Lorentz
equation, also admits a straightforward generalization to
non-Abelian gauge fields.
The extended equation has an infinite number of trajectories as
solutions. The properties of these
paths are shown to impart wave-like properties to the particles in
motion. In view of these results, the motion of a particle is formulated
within the framework of the path integral
formalism which yields a generalized
Schr\"{o}dinger type equation in a general gauge field.
As a further implication of the properties of the trajectories assigned
to a particle, this equation is shown to reduce to a set of equations,
one of them being the Klein-Gordon equation.
\newpage
\begin{center}
{\bf 1. Introduction}
\end{center}

The notion of gauge transformation was introduced by Weyl [1,2]
initially in an attempt to develope a unified theory of
electro-magnetic and gravitational fields. In spite of a lack of
success in achieving the original aim, the formulation has evolved into
a central principle in the description of fields acting on microscopic
particles (for an extensive bibliography, see ref.[3]).
Weyl's proposal provides, essentially, a mechanism to associate an
element of a one dimensional group $\cal G$ with an arbitrary curve
$\rho(AB)$ joining the point $A$ to $B$ in the space-time manifold $\cal M$.
An infinitesimal group element associated with the displacement from
$x$ to $(x+dx)$ is defined by
$U_{(x+dx)x} = (1 + \alpha \phi_\mu dx^\mu)$, where $\phi_\mu$,
$\mu = 0,1,2,3$, are functions on $\cal M$ and $\alpha$ is a
constant. The elements of the type $U_{BA}(\rho)$ associated with
$\rho(AB)$ are computed by repeated multiplications  of the
infinitesimal elements. A scalar function $\Phi_{BA}(\rho)$ may now be
associated with $\rho(AB)$ by
\begin{equation}
    \Phi_{BA}(\rho) = U_{BA}(\rho)\Phi_A
\end{equation}
where $\Phi_A = \Phi_{AA}$ is given. The group elements associated with
infinitesimal closed curves, $\rho_{\!_c}$, define a gauge field
with components $f_{\mu\nu}$ :
\begin{eqnarray}
   U_{ADCBA}(\rho) &=& ( 1 + \alpha f_{\mu\nu} d \sigma^{\mu\nu} )
                                                  \nonumber \\
                   &=& ( 1 + \alpha v_A(d \sigma) )
\end{eqnarray}
where $d \sigma$ is the area enclosed by $\rho_{\!_c}(ABCDA)$.

Although the concepts introduced above are valid with any $\alpha$, Weyl
assumed it to be a real, non-zero constant. This assumption leads to a
non-compact group. Also, $\Phi_{BA}(\rho)$ was interpreted as the length
at $B$ of a `rigid' measuring rod transported along $\rho(AB)$ provided
that the length at $A$ is $\Phi_A$.
The Jacobi identity satisfied by $f_{\mu\nu}$ led
Weyl to conclude that the functions $\phi_\mu$ may be identified with
the electro-magnetic potentials. In an attempt to relate Weyl's
formulation to the formalism of quantum mechanics, London [4]
arbitrarily set $\alpha = i$ in natural units, leading to the
description of electro-magnetism in terms of the group $U(1)$. Weyl
later accepted the $U(1)$ description, and satisfactorily described the
interaction of a charged particle with an electro-magnetic field by
requiring the field-free quantum mechanical equations to be gauge
covariant. This results in replacing $\partial_\mu$ by
$(\partial_\mu - i\phi_\mu)$, the gauge covariant derivative in trivial
coupling with $U(1)$ as fibre. The procedure was later extended to
include non-Abelian fields [5]. This coupling scheme
assumes the availability of the field-free equations which are obtained
by quite independent considerations. In addition to the generalization
of the coupling scheme, Weyl's original construction described by (1)
and (2), has also been extended to include an arbitrary differentiable
manifold $\cal M$ and non-Abelian fields by letting $\phi_\mu = \phi_\mu^l X_l$
where $(\alpha X_l)$, $l=1,2,....,n$, are the generators of a Lie group
$\cal G$ [6]. In an irreducible matrix representation of an
$n$-dimensional $\cal G$, $X_l$ are $n \times n$ matrices.
The identity element will still be denoted by 1.

The motion of a particle coupled to an electro-magnetic field is
described in terms of the gauge-field components $f_{\mu\nu}$, by the
Lorentz equation (see e.g., ref.[7]):
\begin{equation}
   m {\dot{u}}_\mu(s) = f_{\mu\nu}u^\nu
\end{equation}
where $u_\mu = {\dot{x}}_\mu(s)$ and
the dot denotes the derivative with respect to the indicated
argument which in the case of (3) is the arc-length.
The infinitesimal arc-length is given by
$ds^2 = g_{\mu\nu}dx^{\mu}dx^{\nu}$, where
$g_{00} = 1, g_{\mu \nu} = -1, \mu = 1,2,3$, and $g_{\mu \nu} = 0$ otherwise.
Here $x_0 = x^0$ denotes the time.
The analogue of (3) for non-Abelian fields [8] also describes
the motion in terms of the gauge-field components, which in group
theoretical terms, are associated with the infinitesimal elements
corresponding to closed curves as defined by (2). A description in terms
of the global elements should be expected to be more complete as they
contain more information than their infinitesimal constituents.

Motion of a particle has been described in terms of the gauge group
elements for the $n=1$ case [9]. In this scheme, (3) is
interpreted in terms of the group elements associated with a class of
infinitesimal closed curves. While the field components are already
expressed in this manner by (2), a representation of the complete
equation requires an additional construction. With such a representation
available, an extension is obtained by an appropriate replacement of the
elements. In the process, the restriction imposed by the action
principle on the curves becomes redundant which is, therefore, dropped.
Although guided by the
present interpretation of (3), the extension termed the gauge mechanical
principle in itself forms a basic assumption of the formulation. In
this development, the value of $\alpha$ is initially allowed to be an
arbitrary complex constant. However, if the extended equation is to
have any meaningful solutions, then $\alpha$ must be purely imaginary,
providing a derivation of its value. With $\alpha$ purely
imaginary, the equation has an infinite number of trajectories as solutions
termed the `physical paths'. The physical paths that a particle is
allowed to follow are shown to possess
wave-like coherences that impart similar properties to a moving
particle. The wave-like behaviour of particles and the multiplicity of
allowed paths form the basic assumptions of Feynman's path integral
formulation [10,11]. Therefore, the procedure
of the path integral formalism is
used to derive a five-dimensional generalized
Schr\"{o}dinger type equation. This equation was initially conjectured
by St\"{u}ckelberg [12,13] to provide a more satisfactory
description of a relativistic particle with integral spin than the
existing equations. In order to base it on a more solid foundation, the
equation has been derived from a variational principle [14,15].
The present treatment
provides a systematic derivation of this equation [9].
Thus the formulation provides a deductive
formalism for these results which have been considered more satisfactory
on intuitive grounds.

While the above extension is formulated without any restriction on the
curves, not all conceivable trajectories are the solutions of the
resulting equation.
In addition to playing a crucial role in the derivation of
some of the above results, the properties of the solutions
lead to another result that
was previously conjectured, as described below. In addition to accepting
the conjecture of St\"{u}ckelberg, Feynman [16] arbitrarily selected
some periodic solutions of the generalized Schr\"{o}dinger equation in
order to deduce the Klein-Gordon equation. The properties of the
particle paths implied by the present principle lead to a boundary condition
on the five dimensional equation derived by the present methods which in
all other respects  is the same as St\"{u}ckelberg's generalized
Schr\"{o}dinger equation. As a result of the boundary condition, the
solutions of this equation are confined to a collection of periodic
functions. This enables one to decompose the five dimensional equation
into a set of four dimensional equations, one of them being the
Klein-Gordon equation.
These results provide further justification for the use of the extended
principle in studying the motion of particles.

In the present article the original formalism for the $n=1$ case
[9] is generalized to include the non-Abelian gauge fields. While the
generalization is formally quite straightforward, a lack of
commutability poses some technical difficulties, requiring some
improvement in the previous techniques. In Sec.2, the classical
variational principle is expressed in group theoretical terms. This
provides  a generalization of the corresponding result of Vatsya [9]
to include a larger class of Lagrangians and the associated
Euler-Lagrange equations. A representation of (3) is obtained as a special
case. In Sec.3, guided by the form of this representation, the gauge
mechanical principle is formulated for a general gauge field. Relevant
properties of the solutions of the extended equation are studied and
used to describe the motion of a free particle, the double-slit experiment,
and the Aharonov-Bohm effect. In Sec.4, an equation of
motion for a particle coupled to an arbitrary gauge field is derived. In
spite of a loss of commutability, the equation is remarkably similar to
the generalized Schr\"{o}dinger equation with the boundary condition for
the $n=1$ case.
\begin{center}
{\bf 2. The Variational Principle}
\end{center}

In this section, some standard results of the variational calculus are
expressed in group theoretical terms. This leads to a formulation of the
classical action principle and the corresponding Euler-Lagrange
equation that is well suited for its extension.

Let $L'(\dot{x},x,\tau)$ be a Lagrangian defined on curves in
${\cal M}$. For a path $\rho(AB) = x(\tau)$ with $x(\tau_1)=A$,
$x(\tau_2)=B$, the action functional ${S'}_{BA}(\rho) = S'(\tau_1,\tau_2)$
is given by
\begin{equation}
   S'(\tau_1,\tau_2) = \int_{\tau_1}^{\tau_2} L'(\dot{x},x,\tau) d\tau
\end{equation}
Without loss of generality, one may take a point $A'$ as the reference
point and set $x(0) = A'$. By letting $x(\tau) = B'$ a variable point,
one obtains ${S'}_{B'A'}(\rho) = S'(0,\tau)$ for the path $\rho(A',B')$,
which defines a one parameter group with elements
\begin{equation}
   {U'}_{B'A'}(\rho) = {\rm exp}(\alpha {S'}_{B'A'}(\rho)) .
\end{equation}
The element infinitesimally close to the identity
for a displacement $dx$ will be denoted by
$(1 + \alpha v'_A(dx))$, which is obtained by retaining terms up to the
first order in $dx$ in the expansion of $U'_{B'A'}(\rho)$.
Corresponding Weyl's scalar $\Phi'_{BA}$
may now be defined as $\Phi_{BA}$ above,
i.e., $\Phi'_{B'A'}(\rho) = U'_{B'A'}(\rho)\Phi'_{A'}$ where
$\Phi'_{A'}$ is given.

Some conceptual clarity is gained in describing the variational
principle by considering the analogue of $x(\tau)$ in $\cal M'$ obtained
from $\cal M$ by including $\tau$ among the co-ordinate variables
(see e.g., ref[17], ch.1).
Thus
the curve $x(\tau)$ in $\cal M$ corresponds to the set of points
$(x(\tau),\tau)$ in $\cal M'$, eliminating a need for an explicit
reference to the parameterization. A metrical structure on $\cal M'$ is
not needed for this purpose. For convenience, the curves in $\cal M'$
will also be denoted by $\rho$ and $\rho_{\!_c}$ with an indication whenever
needed. It is pertinent to remark here that a curve closed in $\cal M'$
is described by a double valued function $x(\tau)$.
Also, Weyl's construction may be expressed in terms of the curves in
$\cal M'$ by replacing $dx^\mu$ by ${\dot{x}}^\mu d\tau$
without altering the results.

The Lagrangian $L'(\dot{x},x,\tau)$ may be treated as a member of the family
$L'(\dot{x} + \lambda\dot{y}, x + \lambda y, \tau)$ defined on curves
$\rho_{\!_\lambda} = (x(\tau) + \lambda y(\tau))$ where the values $\tau_1$,
$\tau_2$ still correspond to the points $A$, $B$ respectively. This
defines a family of actions
${S'}_{BA}(\rho_{\!_\lambda}) = S'(\tau_1,\tau_2,\lambda)$ obtained by
substituting $L'(\dot{x} + \lambda \dot{y}, x + \lambda y, \tau)$
for $L'(\dot{x},x,\tau)$ in (4).
Let $\rho_{\!_c}(ABA)$ be the union of $\rho_{\!_0}(AB)$
and $\rho_{\!_\lambda}(BA)$. The action
${S'}_{ABA}(\rho_{\!_c})$ on the curve $\rho_{\!_c}$ is given by
\begin{eqnarray}
  {S'}_{ABA}(\rho_{\!_c}) &=& S'(\tau_1,\tau_2,0) + S'(\tau_2,\tau_1,\lambda)
                                                  \nonumber  \\
                   &=& S'(\tau_1,\tau_2,0) - S'(\tau_1,\tau_2,\lambda)
\end{eqnarray}
It is clear that $\rho_{\!_c}$ is closed in $\cal M'$ and hence in $\cal M$.
By definition, the corresponding group element given by
\begin{eqnarray}
   {U'}_{ABA}(\rho_{\!_c}) = {\rm exp}(\alpha {S'}_{ABA}(\rho_{\!_c}))
\end{eqnarray}
is defined in terms of the two parameters, $\tau$ and $\lambda$.
The infinitesimal element $(1+\alpha v'_{BA}(d\sigma))$ is
obtained by retaining terms up to the first order in $\lambda$,
equivalently in $d\sigma$, i.e., $v'_{BA}(d\sigma)$ is the first term in
the expansion of $S'_{ABA}(\rho_{\!_c})$.

The above construction for closed curves is closely related to the
variational principle as follows. It  is clear from (6), that the
deformation of $x(\tau)$ by $\delta x(\tau) = \lambda y(\tau)$ results
in the change $\delta {S'}_{BA}(\rho_{\!_0}) = {S'}_{ABA}(\rho_{\!_c})$
in the action.
Hence, from (7), $\delta {S'}_{BA}(\rho_{\!_0}) = 0$ is equivalent to
$U'_{ABA}(\rho_{\!_c}) = 1$. The variational principle requires the
equality $\delta {S'}_{BA}(\rho_{\!_0}) = 0$
to hold only up to the first order
in $\lambda$, as it varies in an arbitrary but small neighbourhood of
zero, and $y$ is any given, reasonably smooth function. Since
$v'_{BA}(d \sigma) = S'_{ABA}(\rho_{\!_c}) = \delta S'_{BA}(\rho_{\!_0})$
up to the first order in $\lambda$, the variational principle is
equivalent to $v'_{BA}(d \sigma) = 0$ for all curves closed in $\cal M'$
joining the points $(A,\tau_1)$ and $(B,\tau_2)$ with sufficiently small
$d \sigma$.  The curves $\rho_{\!_0} (AB)$ and $\rho_{\!_\lambda} (AB)$
are usually referred to as the
virtual paths and the action associated with the configuration is
significant only to the extent that it yields the classical trajectory.
This construction may be characterized by more elementary
curves, as shown below.

For a given point $x$ on $\rho(AB)$, let $\rho_{\!_c}(x,x+dx,x)$
be a curve closed in $\cal M'$ obtained as $\rho_{\!_c}(ABA)$. As $dx$
tends to zero, it is sufficient to retain terms up to the first order in
$dx$ in defining the corresponding infinitesimal group element
$(1 + \alpha v'_x(d \sigma))$. Since an arbitrary curve
$\rho_{\!_c}(ABA)$ may be expressed as a union of the curves of the type
$\rho_{\!_c}(x,x+dx,x)$, we have that
$\delta S'_{BA}(\rho_{\!_0}) = S'_{ABA}(\rho_{\!_c}) = \sum v_x'(d \sigma)$
where $\sum$ denotes the sum over all of these curves. This is
sufficient to conclude that an element $U_{ABA}'(\rho_{\!_c})$ may be
expressed as a product of infinitesimal elements of the type
$(1 + \alpha v'_x(d \sigma))$. This implies that if $v'_x(d \sigma) = 0$
for each $x$ on $\rho(AB)$, and each $\rho_{\!_c}(x,x+dx,x)$, then
$v'_{BA}(d \sigma) = 0$ for each $\rho_{\!_c}(ABA)$. The converse is
obvious as $\{ \rho_{\!_c}(ABA) \}$ includes the set
$\{ \rho_{\!_c}(x,x+dx,x) \}$. Thus the variational principle is
equivalent to $(1 + \alpha v'_x(d \sigma)) = 1$ as long as the equation
is required to hold for all curves of the type $\rho_{\!_c}(x,x+dx,x)$.

We have shown above that setting $\delta S'_{BA}(\rho_{\!_c}) = 0$ up to
the first order in $\lambda$ with an arbitrary $y(\tau)$ is equivalent to
requiring $(1 + \alpha v'_x(d \sigma)) = 1$ for all arbitrary curves of
the type $\rho_{\!_c}(x,x+dx,x)$. This condition yields the
Euler-Lagrange equation of the variational problem, if it has a
solution. The Euler-Lagrange equation is satisfied if and only if the
action is stationary. The classical action principle assigns the
solution $\rho_{\!_S}(AB)$ of this equation to a particle. The action
$S'_{BA}(\rho_{\!_S}) = S'(A,B,\tau_1,\tau_2)$ along
$\rho_{\!_S}(AB)$, termed
Hamilton's principal function, depends only on the end points
of the curve and the parameter $\tau$. The principal function may be
obtained as a solution of the associated Hamilton-Jacobi equation.
While the elements $U'_{BA}(\rho)$ and $v'_{BA}(\rho)$ may be computed
directly from equations (4) to (7), they may also be constructed in
terms of the action $S'(A,B,\tau_1,\tau_2)$ as follows. In addition to
providing a clear expression for the value of the Lie algebra element
$v'_x(dx)$, quite frequently this construction is more convenient,
especially if only a few terms in the expansion of the group elements
are needed.

Let $\rho_{\!_c}$ be the union  of $\rho_{\!_S}(x,x+dx)$ and
$\rho_{\!_\lambda}(x+dx,x)$ for a small value of $\lambda$ and an
arbitrary $y(\tau)$. Also, let $(1 + \alpha v'_x(d \sigma))$ be the
infinitesimal group element associated with the closed curve
$\rho_{\!_c}$. Since the action $S'(x,x+dx,\tau,\tau + d\tau) = dS'(x,\tau)$
is stationary under the deformation $\rho_{\!_\lambda}(x,x+dx)$ of
$\rho_{\!_S}(x,x+dx)$, equivalently $v'_x(d \sigma) = 0$, we have that
$S'_{x+dx,x}(\rho_{\!_\lambda}) = S(x,x+dx,\tau,\tau + d\tau)$
up to the first order
in $\lambda$. Therefore the infinitesimal group element
$(1 + \alpha v'_x(d x))$ for an arbitrary deformation of
$\rho_{\!_S}(x,x+dx)$ is given by $(1 + \alpha dS'(x,\tau))$ where the
terms only up to the first order in $dx$ have to be retained. Since the
global group element $U'_{BA}(\rho)$ associated with $\rho(AB)$ may be
obtained as a product of elements of
the type $(1 + \alpha v'_x(dx))$, it may be
computed by multiplying the elements $(1 + \alpha dS'(x,\tau))$, yielding
\begin{equation}
  {U'}_{BA}(\rho) = {\rm exp}(\alpha\int_{\rho(AB)} dS'(x,\tau))
\end{equation}
This construction evaluates $S'_{BA}(\rho)$ by expressing it as a sum.
Each term of the sum is the Hamilton's principal function for points $x$
and $(x+dx)$ on $\rho(AB)$. The sum approaches $S'_{BA}(\rho)$ as $dx$
tends to zero.

As a prelude to applying the above results to the Lorentz equation,
consider the classical motion of a free particle which is usually
described by the homogeneous Lagrangian
$L' = L^0(\dot{x},x,s)=m \sqrt{ \dot{x}_\mu(s) \dot{x}^\mu(s) }$.
There is some difficulty with a straightforward application of the
Hamilton-Jacobi construction in case of a homogeneous Lagrangian
(see e.g., ref.[17], ch.3).
One procedure to circumvent the difficulty is to take
$L' = L^P = \frac{1}{2} m (\dot{x}_\mu(\tau)\dot{x}^\mu(\tau) + 1)$ with
$\tau$ an independent parameter and after the construction, set
$\tau=s$, equivalently $\dot{x}_\mu\dot{x}^\mu = 1$ [18,19].
Let the elements $U'_{BA}(\rho)$ and $v'_x(d\sigma)$ obtained by
substituting $L^P$ for $L'$ in (4), (5) and (7) be denoted by
$U_{BA}^P(\rho)$ and $v_x^P(d\sigma)$ respectively. It is clear from the
above that the classical motion of a free particle in a neighbourhood of
$x$ is described by $v_x^P(d\sigma) = 0$ for all closed curves of the type
$\rho_{\!_c}(x,x+dx,x)$.

The Lorentz equation, (3), may be obtained as the Euler-Lagrange
equation of the variational principle with the Lagrangian
$L' = L^0 - \phi_\mu \dot{x}^\mu(s)$ [7].
However, for the same reason as with the free particle, the Lagrangian
$L' = L^P - \phi_\mu \dot{x}^\mu(\tau)$ is preferable [18,19].
With this choice of $L'$ in equations (4) to
(7), the elements are defined by
$U'_{BA}(\rho) = U_{BA}^{-1}(\rho) U_{BA}^P(\rho)$ and
$v'_x(d\sigma) = (v_x^P(d\sigma) - v_x(d\sigma))$, where $U_{BA}(\rho)$ and
$v_x(d\sigma)$ are as defined by (1) and (2). Thus the Lorentz equation is
implied by, and implies, the condition $v'_x(d\sigma) = 0$, i.e.,
$v_x^P(d\sigma) = v_x(d\sigma)$ for all curves of the type
$\rho_{\!_c}(x,x+dx,x)$. This shows that the classical motion of a
charged particle in an electro-magnetic field is characterized by the
equality $(1 + \alpha v_x^P(d \sigma)) = (1 + \alpha v_x(d \sigma))$,
between the infinitesimal group elements. This equality is required to
hold for all curves of the type $\rho_{\!_c}(x,x+dx,x)$.

For later reference, the Weyl's scalar associated with $U_{BA}^P(\rho)$
will be denoted by $\Phi_{BA}^P$. For $n>1$, $U_{BA}^P$ will be assumed
to be multiplied by the identity without an explicit indication and
$\Phi_{BA}^P(\rho)$ will be taken to be an $n$-vector. This makes
$U_{BA}^P(\rho)$ and $\Phi_{BA}^P(\rho)$ compatible with $U_{BA}(\rho)$
and $\Phi_{BA}(\rho)$ respectively. Hamilton's principal function will
not be needed for any Lagrangian other than $L^P$. In the following,
$S(\cdot\cdot\cdot)$ and $dS(\cdot\cdot\cdot)$ will denote this function
and its infinitesimal value, respectively.
\begin{center}
{\bf 3. Physical Paths}
\end{center}

The characterizations of the variational principle and the Lorentz
equation described in Sec.2 indicate that a description of motion
within the framework of the classical action principle is incomplete in
group theoretical terms. One of the inherent limitations is due to the
equality
$(1+\alpha v_x^P(d\sigma)) = (1+\alpha v_x(d\sigma))$
that equates the corresponding global elements only up to the first order in
$d\sigma$. Additional information that may be available in the global
elements is not utilized in the action principle. Also, the
equivalence is required to hold for the curves of the type
$\rho_{\!_c}(x,x+dx,x)$, i.e., only the curves with closed images in
$\cal M'$ which imposes an additional restriction on the solutions.
In this section, we extend the classical treatment by
requiring, essentially, the equivalence of $U_{BA}^P(\rho)$ and
$U_{BA}(\rho)$ and by considering all curves in $\cal M$,
which is a natural extension by completion of the above
equality and thus of the action principle. An explicit introduction of
the parameter is more convenient for some of the following treatment
which will be used accordingly.  With
appropriate construction of the group elements, closed curves are
included in the collection $\{ \rho_{AB} \}$.

{\bf 3.1  The guage mechanical principle. }  A group
element $U_{BA}(\rho)$ for a general gauge field is given by
the path-ordered exponential ${\rm Exp}(\cdot)$ [20]:
\begin{eqnarray}
  U_{BA}(\rho) &=& {\rm Exp} \left( \alpha \int_{\rho(AB)}
                              \phi_\mu(x)dx^\mu \right) \nonumber  \\
               &=& \sum_{l=0}^{\infty} \alpha^l
                  \int_0^1 d\tau_1 \cdot\cdot\cdot
                  \int_0^{\tau_{l-1}} d\tau_l \;\;
                  \left( \right. \dot{\gamma}^{\mu_1}(\tau_1) \cdot\cdot\cdot
                  \dot{\gamma}^{\mu_l}(\tau_l)   \; \nonumber            \\
               & &~~~~~~~\times \;\phi_{\mu_1}(\gamma(\tau_1))
                  \cdot\cdot\cdot
                  \phi_{\mu_l}(\gamma(\tau_l))  \left. \right)
\end{eqnarray}
where $\gamma(\tau_j)$ is a parameterization of
$\rho(AB)$ with $\gamma(0)=A$
and $\gamma(1)=B$.
If $\{X_l\}$ forms a commuting set,
then ${\rm Exp}(\cdot)$ reduces to the ordinary exponential.
The elements $U_{BA}^P (\rho)$ are given by (8):
\begin{equation}
  U_{BA}^P(\rho) = \exp \left( \alpha \int_{\rho(AB)} dS(x,\tau)  \right)
\end{equation}

As explained above, Weyl's vectors $\Phi_{BA}^P(\rho)$ and
$\Phi_{BA}(\rho)$ are defined by $U_{BA}^P(\rho)$ and $U_{BA}(\rho)$
with their values $\Phi_A^P$ and $\Phi_A$ at $A$ being given, by
appropriate substitutions in (1).

Weyl's vectors $\Phi_A$ and $\Phi_A^P$ may be assumed to be related by
$\Phi_A^P = \kappa(A) \Phi_A$ where $\kappa(A)$ is an invertible
matrix depending on the physical properties of the system at $A$. A
particle coupled to a gauge field will be assumed to follow the paths
defined by
\begin{equation}
  \Phi_{BA}^P(\rho) = \kappa(B) \Phi_{BA}(\rho)
\end{equation}
i.e., a path $\rho(AB)$ is allowed (physical) if and only if there are
(physical) points $A$ and $B$ and a trajectory $\rho$ joining them such
that (11) is satisfied.

Eq.(11) is essentially the statement of the gauge mechanical
principle which provides the present extension of the classical
description of motion. Its alternative characterizations and necessary
explainations are given below.

In view of (10), $U_{BA}^P(\rho)$ is a constant multiple of the identity
matrix and hence commutes with all $n \times n$
matrices. Consequently (11) is equivalent to
\begin{equation}
  V_{BA}(\rho) \Phi_A^P = \Phi_A^P
\end{equation}
where $V_{BA}(\rho) = \kappa(B) V_{BA}'(\rho) \kappa^{-1}(A)$ with
$V_{BA}'(\rho) = {[ U_{BA}^P(\rho) ]}^{-1} U_{BA}(\rho)
= U_{AB}^P(\rho)U_{BA}(\rho)$.
Eqs. (11) and (12) may also be expressed as
\begin{equation}
  V_{BA}'(\rho) \Phi_A = \kappa^{-1}(B) \kappa(A) \Phi_A
\end{equation}
As indicated by (12), the present assumption stated in (11), is
essentially a requirement of an equivalence between $V_{BA}(\rho)$ and
the identity element. Similarly (13) is a statement of equivalence
between $V'_{BA}(\rho)$ and the identity. It follows from the
definitions that both of these are equivalent to an equivalence between
$U_{BA}^P(\rho)$ and $U_{BA}(\rho)$.  This equivalence does
not reduce to a strict equality even for the
Abelian case. For the Abelian groups, as is the case with $n=1$,
and for closed curves $\rho_{\!_c}(ABA)$, this does reduce to a strict
equality. Classical action principle was characterized in Sec.2 by a
strict equality for closed curves. However, only an equivalence can be
inferred for general curves from an equality for the closed ones.

Eqs. (12) and (13) are eigenvalue equations, implying that their
solutions, and hence a physical path
is independent of the magnitudes of $\Phi_A^P$ and $\Phi_A$. Therefore,
$\Phi_A^P$ and $\Phi_A$ may be assumed to be normalized in
the complex $n$-vector space $l_2({\cal C}^n)$. This restricts
$\kappa$ to the class of norm-preserving, i.e., unitary matrices.
It will be seen below that a real $n$-vector space is too restrictive to
describe the physical paths. Thus, further
restrictions on $\Phi_A$, $\Phi_A^P$ and $\kappa$ are not possible.
The matrix $\kappa(A)$ embodies the properties of the physical system
under consideration at $A$, which should remain unchanged if $\phi_\mu = 0$.
Therefore it will be assumed that if $\phi_\mu = 0$, for each $\mu$ and
all paths in a set large enough to include
physical paths joining $A$ and $B$, then $\kappa(A) = \kappa(B)$.

The constant $\alpha$ in the above is as yet undetermined. Its value was
determined in ref.[9] by considering the motion of a charged
particle in an electro-magnetic field. The same result is obtained by
considering the motion of a free particle, as follows.

For a free particle, i.e., in the limit of $\phi_\mu = 0$, the physical
paths are defined by
\begin{equation}
 \exp \left( -\alpha S_{BA}(\rho) \right) \Phi_A^P = \Phi_A^P
\end{equation}

Since $S_{BA}(\rho)$ is in general real, it follows that $\alpha$ must
be purely imaginary.
The trivial solution $\alpha = 0$ is excluded for it implies that neither
the gauge field nor the geometry of ${\cal M}$ has any effect on the
motion.

Present formulation does not yield the magnitude of $\alpha$ which must
be determined experimentally. The elements of the type $U_{BA}(\rho)$
with an imaginary value of $\alpha$ are called the phase-factors
and the resulting group is compact. By comparing with the usual
definition of the phase-factors that conforms with the experimental
observations, $\alpha = 2 \pi i /h$ where  $h$ is Planck's constant.
Thus, $\alpha$ may be set equal to $i$ by selecting units with
$h = 2 \pi$.
This identification will be seen to be consistent with the rest of the
treatment to follow. Natural units are used elsewhere also.

Solutions of (12) are identified by their equivalence classes as
follows.  Let $\{B_j\}$ be a set of points on $\rho(AB)$
such that $V_{B_jA}(\rho)\Phi_A^P = \Phi_A^P$.  If one member
of $\{B_j\}$ is a physical point with respect to $\{ \rho(AB),\Phi_A^P \}$
, then this is also the case
for each $j$.  Thus the equivalence class $\{B_j\}$ so defined
characterizes the
solutions $\{\rho(AB_j)\}$. A natural order is defined on $\{B_j\}$
by setting  $B_j$ to be the $j$th
closest member to $A$.  Let $\{B_j^k\}$, $k=1,2,....$,
be such ordered equivalence classes with respect to
$\{\rho(AB^k),\Phi_A^P\}$.
The set $\zeta_j = \{B_j^k \}$  defines a physical `surface' for each $j$.

For a free particle, the physical paths are the
solutions of (14) which reduces to
\begin{equation}
\exp \left( -im \int_{\rho(AB)} u_\mu dx^\mu \right) \Phi_A^P = \Phi_A^P
\end{equation}
The equivalent points
$\{B_j\}$ on these curves satisfy
\[  m \int_{\rho(B_j,B_{j+1})} u_\mu dx^\mu = 2 \pi    \]
Along the paths characterized by a constant velocity $\bar{u}$, $B_j$ and
$B_{j+1}$ are thus separated by the de
Broglie wavelength $2 \pi / m \bar{u}$ and the length of a physical path
is its integral multiple.

Consider a source-detector system with source at $A$ and detector at $B$.
A curve $\rho (AB)$ will be called monotonic if the parameter value increases
or decreases monotonically along the curve.  By convention, $\tau$ will be
assumed to increase from $A$ to $B$.  A particle starting at $A$ and confined
to $\rho (AB)$ is observable at $B$ if and only if $\rho (AB)$ is physical.
If $\theta$  is the intensity associated with
$\rho (AB)$ at $A$ then the intensity transmitted
to $B$ by this path must be equal to $\theta$.

A union of physical paths is obviously physical.  Also a union of
non-physical monotonic curves can be physical.  For example, let $\rho (AB)$
be a monotonic physical path with the associated physical points $\{ B_j \}$
and let $C$ be a point in the interior of $\rho(B_j B_{j+1})$.
Then the union of $\rho (B_j C)$ and $\rho (CB_{j+1})$ is
$\rho (B_{j} B_{j+1})$
and the union of $\rho(B_{j-1} C)$ and $\rho (CB_j)$ is
$\rho (B_{j-1} B_{j})$,
both of which are physical.  However, these trivial constructions are
redundant as they are indistinguishable from the paths of the type
$\rho(B_k B_{k+l})$.
A significant, non-trivial class of such paths is described below.

Consider a configuration of two curves
$\rho(AB)$ and  $\rho'(AB)$ with  $\rho_{\!_c}(ABA)$
being the union of $\rho'(AB)$ and $\rho(BA)$,
i.e., the type of the virtual paths
encountered in the action principle.  In classical mechanics, these
curves serve only to define the classical trajectory.  According to the
present prescription, if (11) is satisfied then this is a physical
configuration.  In that case, $\rho(AB)$ and $\rho'(AB)$ offer equally likely
alternatives for the transmission of a particle from $A$ to $B$, even if
$\rho(AB)$ and $\rho'(AB)$ may not be physical.
The case of the alternatives of
the type $\rho(AB)$ and $\rho'(CB)$ is treated similarly.
To be precise, let the parameter value at B be $\tau_{\!_B}$.
According to the above convention, $\tau$
increases from $C$ to $B$ along $\rho'(CB)$ and decreases from
$B$ to $A$ along $\rho(BA)$.
The group elements associated with such configurations may be
computed by integrating along  $\rho'(CB)$ and then along $\rho(BA)$.
If $V_{ABA} (\rho'') \Phi_A^P = \Phi_A^P$,
where $\rho''$ is the union of  $\rho'(CB)$ and  $\rho(BA)$,
then  $\rho(AB)$ and  $\rho'(CB)$ offer
likely alternatives.  Such configurations of trajectories are referred
to as the interfering alternatives.  The intensity of particles
transmitted to $B$ by the equally likely alternatives must be equal to
the sum of the intensities at $A$ and $C$ associated with the respective
trajectories.  A separate treatment of such configurations is not
necessary for a general theory as they are represented by well
defined curves in $\cal M'$ , but it provides clearer understanding of some
physical phenomena described in the sequel.

As a prelude to a precise treatment of motion in Sec.4, an approximate
description of a few phenomena is given in Secs.3.2, 3.3 and 3.4,
which also clarifies the properties of a multiplicity of physical
trajectories.

{\bf 3.2  Motion of a free particle.}  Consider a physical system described by
a Lagrangian $L'(\dot{x},x)$ with $\rho_{\!_S}$ being
the resulting classical path.
For convenience, it is assumed that $L'$ does not depend on $\tau$ explicitly.
However, $\tau$-dependence may be included without a significant change in
the following analysis.  For a free particle, $L' = L^p$.
For an undisturbed particle,  the equivalent points on $\rho_{\!_S}$
are given by
\[  S'(B_j,B_{j+1},\tau_j,\tau_{j+1}) = 2 \pi  \]
where $S'( \hspace{.15in}  )$ denotes Hamilton's principal function.

The action $S'_{B'A'}(\rho')$ along a trajectory  $\rho'(A'B)$
in a small neighbourhood
of  $\rho(AB)$ may be expressed as
\begin{eqnarray}
 \left( S'_{B'A'}(\rho')-S'_{BA}(\rho) \right) &=&
 \int_{\rho(AB)} \delta x^\mu \left[ \frac{ \partial L'}{\partial x^\mu}
 - \frac{d}{d \tau} \frac{ \partial L'}{ \partial \dot{x}^\mu} \right] d\tau
   \nonumber \\
                                               & &
   + \left[ \frac{\partial L'}{\partial \dot{x}^\mu} \delta' x^\mu
          - H \delta' \tau \right]_A^B + O(\delta^2)
\end{eqnarray}
by standard methods [21].  Here $\delta' x^\mu$, $\delta' \tau$
correspond to the variation of the
end points $A$, $B$ to $A'$, $B'$, and $H$ is the Hamiltonian.
The term $O(\delta^2)$ is
the integral along  $\rho(AB)$ of $\cal J$ , containing functions of
second or higher
order in $(\delta x)$ and $(\delta \dot{x})$.

If $\rho = \rho_{\!_S}$ , then the first term on the right side of (16) is
equal to zero.
Hence $S'_{B'A'} (\rho') = S'_{BA} (\rho_{\!_S})$ for some values of
$\delta' x = O(\delta^2 )$.  Therefore the
trajectories in a $\delta x$ neighbourhood of a physical
classical path  $\rho_{\!_S}(B_j B_{j+k})$
are also physical and their end points are confined to $(\delta^2)$
neighbourhoods of $B_j$ and $B_{j+k}$.  Thus the intensity transmitted
by paths in
a $\delta x$ neighbourhood of a classical trajectory
is concentrated in $(\delta^2)$
neighbourhoods of the equivalent points on $\rho_{\!_S}$.
Let $\rho$ be a path
transmitting intensity outside $(\delta^2)$ neighbourhood of $\{ B_j \}$,
i.e. $\rho$ is outside
$(\delta x)$ neighbourhood of $\rho_{\!_S}$.  Since $\rho$ is not a solution
of the Euler-Lagrange
equation, the first term in (16) dominates which is $O(\delta x)$.
Repeating the
above argument, we have that the intensity transmitted by trajectories
in a $\delta x$ neighbourhood of $\rho$ is spread over a $\delta x$
neighbourhood of points
outside $(\delta^2)$ neighbourhood of $\{ B_j \}$.  Further, the magnitude
of the first
term in (16) increases as $\rho$ is removed farther from the classical
trajectory.  Therefore the contribution to the intensity decreases
accordingly.  Some intensity is also transmitted by the interfering
alternatives whose monotonic segments are non-physical.  In a
homogeneous space, such paths are roughly evenly distributed about the
classical trajectory implying a uniform distribution of the associated
intensity.  The properties of these paths will be described in more
detail in Sec.3.3 where their impact is greater.

Assuming that the particles originate in a small region about a point $A$,
intensity should be expected to be higher near the points equivalent to
$A$ and to decrease away from them, creating a wave-like pattern over a
uniform background.  On a classical scale, the segments between
$B_j$ and $B_{j+1}$
are negligibly small.  Also for macroscopic trajectories, the
contribution of the first term in (16) is enormous as one moves away
from a purely classical trajectory, owing to the large interval of
integration.  Therefore, the contribution to the variation of the
intensity over a wavelength, between $B_j$ and $B_{j+1}$ , must
come from extremely
small neighbourhoods of the the long trajectories, and from larger
neighbourhoods of the shorter ones, which are still small on a classical
scale.  Thus on a macroscopic scale, the particles from $A$ to $B$ travel
along narrow beams centered about the classical trajectories.  A
classical path has an uncertainty associated with it to this extent.  In
view of the above, the intensity should be concentrated in a small
neighbourhood of $B$ decreasing away from the central point.

{\bf 3.3  The double-slit experiment.}  The interfering alternatives play a
prominent role in the double slit experiment.  The this setup, identical
particles are allowed to pass through two slits at $A$ and $A'$, and
collected on a distant screen at a point $B$.  As explained in Sec.3.2,
the particle paths may be assumed concentrated about the classical
trajectories from $A$ to $B$ and from $A'$ to $B$.  If one of the beams is
blocked, then the intensity observed in a neighbourhood of $B$ should
behave almost as deduced in Sec.3.2 for a free particle.  However, if
the intensity is transmitted by both of the beams, then a multitude of
the interfering alternatives is allowed.  Existence of such paths and
their influence on the intensity distribution is studied next.

In view of the physical equivalence of A and A' and that of the
particles, one has that  $\kappa(A) = \kappa(A')$,
$\Phi_A^P = \Phi_{A'}^P$, and hence $\Phi_A = \Phi_{A'}$.
However, because of
an interaction with the detecting instrument at $B$, $\kappa(B)$
may not be equal
to $\kappa(A)$. For monotonic trajectories, this dislocates only one physical
point about $B$, having no significant impact on the above conclusions.
For the interfering alternatives, $\kappa(B)$ cancels out and the paths are the
solutions of
\begin{equation}
      exp \left[ i \left( \int_{\rho(AB)} dS(x,\tau) -
             \int_{\rho'(A'B)} dS(x,\tau) \right) \right]\Phi_A = \Phi_A
\end{equation}
For the classical trajectories, $\rho = \rho_{\!_S}$
and $\rho' = \rho'_{\!_S}$,
(17) is solved by
\[
    \left( S'_{BA} (\rho_{\!_S}) - S'_{BA'} (\rho'_{\!_S}) \right) = 2 \pi j
\]
where $j$ is an arbitrary integer and the action in this case is
Hamilton's principal function or the arc-length in $\cal M$.  Classical paths
are characterized by a constant velocity $\bar{u}$.  This reduces the solution
to $\Delta r = 2 \pi j/m \bar{u}$, where $\Delta r$ is the difference
between the path-lengths of
$\rho_{\!_S}(AB)$ and $\rho'_{\!_S}(A'B)$.  Therefore $\rho_{\!_S}(AB)$ and
$\rho'_{\!_S}(A'B)$ are interfering alternatives
whenever $\Delta r = 2 \pi j/m \bar{u}$.

Let $B(\varepsilon)$ be the point on the screen such that
\begin{equation}
 \left( (S'_{B(\varepsilon)A} (\rho_{\!_S}) -
         S'_{B(\varepsilon)A'} (\rho'_{\!_S}) \right) = 2 \pi (j + \varepsilon)
\end{equation}
for a fixed $j$ and each $0 \leq \varepsilon \leq  1/2$.
In the following we study the
variation of intensity as $\varepsilon$ varies in the prescribed
interval which is
sufficient to describe it on the entire screen.

It follows from the analysis of Sec.3.2, that
$S'_{CA} (\rho) = S'_{B(\varepsilon)A} (\rho_{\!_S})$,
$S'_{C'A'} (\rho') = S'_{B(\varepsilon)A'} (\rho'_{\!_S})$,
for $\rho$, $\rho'$ in $\delta x$ neighbourhoods of
$\rho_{\!_S}$, $\rho'_{\!_S}$
respectively, where $C$ and $C'$ vary
over a $(\delta^2)$ neighbourhood of $B(\varepsilon)$ on the screen for
a fixed $\varepsilon$.  Therefore,
by varying the paths over a $(\delta x)$ width of the beam and over a
$(\delta^2)$
neighbourhood of $B(\varepsilon)$ it is possible to satisfy
\[
  \left( S'_{DA} (\rho) - S'_{DA'}(\rho') \right) = 2 \pi (j + \varepsilon)
\]
for most of the paths.  In fact cancellations favour this equality which
can be easily seen, in particular for the cases when $\rho_{\!_S}$,
$\rho'_{\!_S}$ are extremals as
is presently the case.  This conclusion is valid for other points in the
vicinity of $A$ and $A'$ also.  For $\varepsilon = 0$, this implies that
there is a large
concentration of interfering alternatives reaching about $B(0)$ and hence
the intensity in a $(\delta^2)$ neighbourhood of $B(0)$ is almost equal to the
intensity in $\delta x$ neighbourhoods of $\rho_{\!_S}(AB(0))$ and
$\rho'_{\!_S}(A'B(0))$.  For $\varepsilon \neq 0$, the
configuration of the paths $\rho_{\!_S}(AB(\varepsilon))$ and
$\rho'_{\!_S}(A'B(\varepsilon))$ is obviously non-physical. From the above
argument, a large number of paths in $\delta x$ neighbourhoods of
$\rho_{\!_S}(AB(\varepsilon))$ and $\rho'_{\!_S}(A'B(\varepsilon))$
are excluded from
combining to form the interfering alternatives and hence unable to
transmit the intensity in a $(\delta^2)$ neighbourhood of
$B(\varepsilon)$.  However, still
there are many paths capable of transmitting intensity about
$B(\varepsilon)$ for $\varepsilon \neq 0$,
which are described below.

It follows from (16) that for trajectories $\rho(AB(\varepsilon))$,
$\rho'(A'B(\varepsilon))$ in $\delta x$
neighbourhoods of $\rho_{\!_S}(AB(\varepsilon))$,
$\rho'_{\!_S}(A'B(\varepsilon))$ respectively,
\[
  (S'_{B(\varepsilon)A}(\rho) - S'_{B(\varepsilon)A}(\rho_{\!_S}))
  = O(\delta^2)
\]
and
\[
  (S'_{B(\varepsilon)A'}(\rho') - S'_{B(\varepsilon)A'}(\rho'_{\!_S}))
  = O(\delta^2)
\]
We have used the fact that the first term on the right side of (16) is
zero as the curves are varied about the classical trajectories and the
second term is zero as the end points are kept fixed.  For these curves,
we have
\begin{equation}
    (S'_{B(\varepsilon)A}(\rho) - S'_{B(\varepsilon)A'}(\rho'))
  = 2 \pi (j + \varepsilon) + O(\delta^2)
\end{equation}
Since there are distortions for which $O(\delta^2)$ term is non-zero and its
magnitude is large in natural units, it is possible to adjust the curves
$\rho$, $\rho'$   such that
\begin{equation}
   (S'_{B(\varepsilon)A}(\rho) - S'_{B(\varepsilon)A'}(\rho')) = 2 \pi k
\end{equation}
with $k = j$ or $(j+1)$, most likely $j$.  This implies that
$\rho(AB(\varepsilon))$ and
$\rho'(A'B(\varepsilon))$ form a pair of interfering alternatives.
Since $\rho(AB(\varepsilon))$, $\rho'(A'B(\varepsilon))$
are non-classical trajectories, it follows as in Sec.3.2 that while
there is a multitude of paths satisfying (20), in $\delta x$ neighbourhoods of
the central paths, their end points are spread over a $\delta x(\varepsilon)$
neighbourhood
of $B(\varepsilon)$.  This implies that the amount of intensity that is
concentrated
in a $(\delta^2)$ neighbourhood of $B(0)$ is spread over a
$\delta x(\varepsilon)$ neighbourhood of $B(\varepsilon)$.
Consequently, a rapid decrease in the intensity is expected as $\varepsilon$
increases away from zero.

As $\varepsilon$ increases further, it is seen from (19) that the
neighbourhood $\delta x$ must
be increased to satisfy (20), i.e. $\rho(AB(\varepsilon))$,
$\rho'(A'B(\varepsilon))$ must be moved farther
away from the solutions of the Euler-Lagrange equations.  Thus the
magnitude of the first term on the right side of (16) integrated along
$\rho(AB(\varepsilon))$, $\rho'(A'B(\varepsilon))$ increases as $\varepsilon$
increases for each fixed variation $\delta x$.  As
above, $O(\delta x(\varepsilon))$ increases with $\varepsilon$,
implying a decrease in the intensity.

The above arguments also imply a symmetric intensity distribution as
$\varepsilon$ is
varied over the interval zero to -1/2, and a repeat of the pattern as $j$
is varied over the integers.  Thus an interference pattern should be
observed on the screen over a background of almost uniform but
relatively low intensity as the major contributions have been estimated here.

Similar arguments may be used to estimate the variations in the
intensity about peaks as $j$ varies, resulting in a decrease in the
intensity as $\mid j \mid$ increases.
This result is based on the fact that the
term $O(\delta^2)$ for each $j$, may be expressed as a sum of two terms,
one being
$j$-independent and the other, directly proportional to $\mid j \mid$.

Availability of two interfering beams originating at $A$, $A'$ and the
equivalence of the physical conditions at these points have played a
crucial role in the above analysis.  As explained before, if one of the
beams is blocked, the interference pattern is destroyed.  Also, such a
distribution should not be expected to result if the equivalence of $A$
and $A'$ is violated.

Above considerations indicate a wave-like behaviour of microscopic
particles observed macroscopically as a collection while behaving as
particles individually.  This is in agreement with the observed
behaviour [11, pp. 2-5].  These results obtained here from (11), are
known to inspire the formalism of quantum mechanics.

{\bf 3.4  The Aharonov-Bohm effect.}   Additional insight into the behaviour
of the particles as implied by the present extension may be gained by
considering their response to a non-zero guage field, as follows.  The
phase-factor obrained from (9) by replacing $\phi_\mu$ by $\hat{\phi}_\mu$
will be denoted by $\hat{U}_{BA}(\rho)$.
Let $\{ \rho \}$, $\{ \hat{\rho} \}$ be the collections of the solutions
of (11), equivalently,
the solutions of (12) with $\phi_\mu$, $\hat{\phi}_\mu$ respectively.
Assume that $U_{BA}(\rho) \neq \hat{U}_{BA}(\rho)$ for a
solution $\rho(AB)$.  If $U_{BA} (\rho)$ is replaced by $\hat{U}_{BA}(\rho)$
in (11), then $\rho(AB)$ is no
longer a solution.  The same conclusion holds for a path $\hat{\rho}(A'B')$.
Thus,
if the inequality holds for some of the solutions of (11) with
$\phi_\mu$, or with $\hat{\phi}_\mu$, then the collections $\{ \rho \}$,
and $\{ \hat{\rho} \}$
of the physical paths are not identical.
Therefore a change of potentials from $\phi_\mu$ to $\hat{\phi}_\mu$
should in general produce an
observable effect.  However, if $U_{BA}(\rho') = \hat{U}_{BA}(\rho')$
for each $\rho'(AB)$ in a
collection $\{ \rho' \}$ large enough to include the union of
$\{ \rho \}$ and $\{ \hat{\rho} \}$, then (12)
remains the same equation under the change from $\phi_\mu$ to
$\hat{\phi}_\mu$.  Consequently, a
change of potential from $\phi_\mu$ to
$\hat{\phi}_\mu$  would not change the solutions $\{ \rho \}$.  Since
the set of physical paths remains the same under this change, the
response of the particles must remain unchanged also.  Therefore, such a
change of potentials will not alter the outcome of an experimental observation.

As an application, consider the Aharonov-Bohm effect [22].  In the
corresponding experimental set up, the electrons travel in beams
centered about paths $\rho(ACB)$ and $\rho'(ADB)$, enclosing a non-zero
magnetic
field but shielded from it.  Chambers used reflectors at $C$ and $D$ to
obtain a configuration of piece-wise classical narrow beams centered
about $\rho(AC)$, $\rho(CB)$, $\rho'(AD)$ and $\rho'(DB)$ [23].
The magnetic field was generated
by placing a long coil carrying an electric current between the
reflectors and perpendicular to the plane of the beams with one end in
the plane.  The electron beams were further shielded from the magnetic
field.  As the current in the coil is varied, the magnietic field varies
accordingly.  The classical Lagrangian for this system is the same as
for the Lorentz equation.

As in the case of the double slit experiment, most of the electrons are
transmitted by the interfering alternatives with parameter value
increasing from $A$ to $B$ along $\rho(ACB)$ and decreasing from $B$
to $A$ along
$\rho'(BDA)$, taking value $\tau_{\!_B}$ at $B$.  The estimates obtained
in the treatment of
the double slit experiment are valid for the present case as they were
not restricted to a free particle.  Some consideration should be given to
the reflectors at $C$ and $D$.  Because of the continuity of the physical
paths at points about $A$, $B$, $C$, and $D$,
$\kappa( \hspace{.15in} )$ cancels out.
{}From Sec.3.2, we
have that most of the intensity transmitted along $\rho(AC)$ reaches a small
neighbourhood of $C$ which remains almost within a macroscopically narrow
beam.  By the same argument, most of this intensity reaches a small
neighbourhood of $B$.  The same comment is valid for $\rho'(ADB)$.
The intensity
along both of the beams is assumed equal.  Consequently, the arguments
of Sec.3.3 can be used to conclude the existence of a similar
interference pattern on the screen.

It follows from (12) that the interfering alternatives for an
electro-magnetic potential $\phi_\mu$ are the solutions of (21):
\begin{equation}
 \exp \left[ -i \oint (dS(x,\tau) - \phi_\mu dx^\mu) \right] \Phi_A = \Phi_A
\end{equation}
where the integration is along the closed curves $\rho_{\!_c}(ACBDA)$.
Here the
phase-factor $U_{BA}(\rho)$ is given by
\[
 U_{BA}(\rho) = \exp ( i \int_{\rho(AB)} \phi_\mu dx^\mu )
\]
It is clear that $\phi_\mu$-dependent part in (21) is $U_{ABA}(\rho_{\!_c})$
which is given by $U_{ABA}(\rho_{\!_c}) = \exp (i F(\phi))$
where $F(\phi)$ is the magnetic flux enclosed by $\rho_{\!_c}$.
As $\rho_{\!_c}$ is
distorted, $F(\phi)$ remains unchanged as long as the distorted closed path
encloses the flux, which covers all of the paths of significance here
as all of them surround the coil.

As $F(\phi)$ varies to $F(\hat{\phi})$,
$U_{ABA}(\rho_{\!_c}) \neq  \hat{U}_{ABA}(\rho_{\!_c})$ for any
$\rho_{\!_c}$  unless
\begin{eqnarray}
(F(\phi) - F(\hat{\phi})) & = & \oint (\phi_\mu
                               - \hat{\phi}_\mu)dx^\mu \nonumber \\
                          & = & 2 \pi j
\end{eqnarray}
with an arbitrary integer $j$.  When (22) is satisfied,
$U_{ABA}(\rho_{\!_c}) = \hat{U}_{ABA}(\rho_{\!_c})$ for
each curve $\rho_{\!_c}$ and hence the experimental observation with
$\hat{\phi}_\mu$ must be the same
as with $\phi_\mu$.  Thus the interference pattern on the screen should repeat
itself periodically as the potential is varied continuously.  The period
is defined by (22).

Let $\phi_\mu(\varepsilon)$ be a one parameter family of potentials with
$0 \leq \varepsilon \leq 1$, such that
$(F(\phi(\varepsilon)) - F(\phi(0))) = 2 \pi$ , i.e., $\varepsilon$
covers one period.  The interference
patterns corresponding to $\phi_\mu(0)$ and $\phi_\mu(1)$ are
indistinguishable.  Let the
solutions of (21) with $\phi_\mu$ replaced by $\phi_\mu(\varepsilon)$
be $\{ \rho(\varepsilon) \}$.  Owing to the continuity
of $F(\phi(\varepsilon))$ with respect to $\varepsilon$,
$\{ \rho(\varepsilon) \}$ should vary continuously, implying a
continuous variation of the corresponding interference pattern.  As
$\varepsilon$ approaches one, the distribution of the intensity
must return to the
same as for $\varepsilon = 0$.  Thus, each interference fringe
should be expected to
shift as $\varepsilon$ varies from zero to one, from its position to
the original location of the next.

Above conclusion agrees with the experimental observation [23,24].  It
is pertinent to remark that the indistinguishability of
$\phi_\mu$ and $\hat{\phi}_\mu$ that satisfy
(22) is a direct consequence of (21) which is obtained from (11) and the
fact that the physical paths in this case are closed in $\cal M'$.
For this part
of the conclusion, no estimates are needed.  Dirac's quantization of the
magnetic monopole is also derivable from (22) [25].

The Aharanov-Bohm effect is an implication of the quantum mechanical
equations [22] which were developed from different premises than the
present formalism.  Ingredients of the quantum mechanical deduction of
this effect are the representation of
the momenta $p_\mu$ by $-i \partial_\mu$ and the
corresponding extension of the classical coupling scheme
$( p_\mu - \phi_\mu)$.  The
former was inspired by the observed wave-like behaviour of particles and
the later, in addition to being intuitive, sets $\alpha= i$ in the London-Weyl
description of electro-magnetism.  Here the major aspects of the
Aharonov-Bohm effect are deduced directly from (11) without an appeal to
any other theory.

\begin{center}
{\bf 4. Equation of Motion}
\end{center}

As explained in Sec. 2, the action principle includes a collection of
curves into consideration, but assigns a unique trajectory to a particle
in motion between two points.  The present extension (11), on the other
hand, assigns many paths, but not all curves are allowed.  Since it is
impossible to assign a unique trajectory to a particle, as an
alternative, one may describe its motion in terms of the intensity of
the particles transmitted to a region in $\cal M$ or $\cal M'$
by the physical
trajectories.  Equivalently, a probability density may be assigned to a
given region and compute it from the contributions from all of the
physical paths passing through the region.  This was done in Sec. 3 for
a beam of free particles and for the double-slit experiment, but only
approximately.  Approximations were made in obtaining the estimates and
by retaining only the major contributions.  In a complete theory, all
physical paths must be included and the contributions must be computed exactly.

Wave-like behaviour of particles and a possibility of describing their
motion in terms of the probability densities associated with a
collection of trajectories led Feynman to develop his path integral
formulation of non-relativistic quantum mechanics [10,11].  The
formalism was extended in an analogous manner by introducing a proper
time-like evolution parameter [16].  The wave-like behaviour of the
particles was used to conclude that the intensity is the absolute square
of the amplitude obtained by the law of superposition.  The amplitude
associated with a path $\rho(AB)$ was taken to be proportional
to $\exp ( i S'_{BA}(\rho))$
which was based on a deduction by Dirac [26] of the behaviour of a
quantum mechanical particle.  This treatment also implies a multiplicity
of paths for a particle.  Present formulation associates a phase-factor
equal to $\exp ( i S'_{BA}(\rho))$ with $\rho(AB)$ whenever a
classical description is
possible in terms of a Lagrangian.  In the process, an alternative
method to compute the phase-factors up to the required order is
developed that has some advantages over the standard procedure [11].
The phases associated with a multiplicity of paths are shown to
interfere in a manner that imparts wave-like properties to the
particles in motion.  A precise determination of a multitude of physical
trajectories follows from (11).  Thus all of the necessary assumptions
required for the formulation of Feynman's postulates have been deduced
from (11).  Having yielded its basic assumptions, the guage mechanical
principle finds a natural expression within the framework of the path
integral formalism.  However, only the physical paths should be included
in the computation of the total contribution.

Above deduction is based on a treatment of some physical systems that
can be described classically within the framework of the action
principle. This includes the response of a charged particle to an
electro-magnetic field.  However, the extension stated in (11) is not
subject to this restriction which is, therefore, dropped in the
following.  Also, the original postulates formulated for $n = 1$ [9] are
extendable in a standard manner to include the non-Abelian gauge fields,
as stated below.
\\[2ex]
{\em Postulate 1}.~~~The probability of finding a particle in
a region of space-time is the square of the $l_2({\cal C}^n)$-norm of the
sum of contributions from each physical path or its segment in the region.
\\[1.5ex]
{\em Postulate 2}.~~~The contribution at a point $C$ of a physical path
$\rho(AB)$ is equal to
$K \kappa^{-1}(C) V_{CA}(\rho) \Phi_A^P = K V_{CA}'(\rho) \Phi_A$
where $K$ is a path-independent constant.  \\[.6ex]

Since the assumptions underlying the above postulates are deduced from
(11), the formatlism is self-consistent and based essentially on one
assumption.  Postulate 2. provides a mechanism for an accurate
computation of the total contribution from all trajectories by the
techniques developed originally for the path-integral formulation.  An
equation of motion is developed below by this procedure and by isolating
the contribution of the physical paths.  Postulate 1. provides a means
to obtain experimentally observable quantities from the solutions of the
equation of motion.

Consider a point $C$ on a physical path $\rho(A'B')$. Let $\rho(AB)$ be
the shortest segment of $\rho(A'B')$ containing $C$ such that $A$ and
$B$ are equivalent to $A'$ and $B'$ respectively. Consider the pair of
points $A$ and $A'$. The pair $B$, $B'$ is treated similarly. In view of
the equivalence, $V_{AA'}(\rho)\Phi_{A'}^P = \Phi_{A'}^P = \Phi_{A}^P$,
we have that
$V_{CA'}(\rho)\Phi_{A'}^P = V_{CA}(\rho)\Phi_{A}^P$. Thus
the contribution from $\rho(A'C)$ is the same as that from $\rho(AC)$.
Therefore it is sufficient to consider the minimal curves $\rho(AB)$
instead of any larger physical paths containing $\rho(AB)$.

The next step is to parametrize the minimal physical paths in a way
that enables one to isolate their contribution.
Since a single parameter is needed for all of the curves, standard
parametrization by arc-length is inadequate. A suitable parameter was
found in ref.[9] as follows. Let $u'_\mu = \sum u_\mu$ where
$\sum$ denotes the sum over all
paths of the type $\rho(AB)$ with $A$ being a variable point. For any
such collection of curves, there is a
Lorentz frame $\cal L$ in which $u_\mu' = 0$ for $\mu=1,2,3$. A particle
may thus be treated as being located at the origin of $\cal L$. Incidentally,
the origin of $\cal L$ coincides with the centre of mass of a fluid of
uniform density and total mass $m$ with an infinitesimal element
flowing along each of $\rho(AB)$. Let $z(\tau)$ be a
parameterization of each path $\rho(AB)$ with $z(0)=A$, where $\tau$ is
the proper time of $\cal L$. In $\cal L$, each of the curves $\rho(AB)$
coincides with the straight line along $\tau$. Therefore,
$V_{CA}(\rho) = \exp(im\tau)$ and hence $B = z(2\pi/m)$. From Postulate
2, the contribution $\psi'(x,\tau)$ at $C = x$ is given by
\begin{equation}
  \psi'(x,\tau) = \sum K' V'[x,z(\tau)] \Phi[z(0)]
\end{equation}
where the sum is over all paths passing through $x$ at $\tau$;
$\Phi [z(0)] = \Phi_A$ and for each
$z(\tau)$, $V'[x,z(\tau)] = V_{CA}'[z(\tau)]$.
The sum is the limit of a finite one with constant $K'$ depending on the
number of terms. Because of the continuity of the paths,
the number of curves for
$\tau=0$ is the same as for $\tau=2\pi/m$. Also, for each physical path
$z(\tau)$, $V[x,z(0)] = V[x,z(2\pi/m)] = 1$, i.e.,
$V'[x,z(0)] = V'[x,z(2\pi/m)] = \kappa^{-1}(C) \kappa(A)$. It follows
that
\begin{equation}
   \psi'(x,0) = \psi'(x,2\pi/m)
\end{equation}

The boundary condition given by [24] provides a means to retain the
contribution in (23) from the physical paths. Thus the proper time $\tau$ of
$\cal L$ acquires a physical significance, which
is treated below as an independent parameter as in [18] and [19]. The
following derivation is essentially the same as in the standard path
integral formulation, except for some technicalities arising out of the
non-Abelian nature of the fields under consideration.

Let $[0,2\pi/m]$ be divided into $N$ equal intervals $[\tau_j,\tau_{j+1}]$,
$j=0,1,...,\mbox{$N-1$}$; with $\tau_0 = 0$, and $\tau_N = 2\pi/m$.
Consider all
of the paths with $z(\tau_k) = {(x)}_k$. By the same argument as for
$n=1$, the vector $\psi'[{(x)}_k,\tau_k]$, for each $k$, is given by
\begin{eqnarray}
  \psi'[(x)_k,\tau_k] &=&
    \int U^P[{(x)}_0,{(x)}_1] \cdot\cdot\cdot U^P[{(x)}_{k-1},{(x)}_k]
    \nonumber \\
 && ~~~~~\times ~ U[{(x)}_k,{(x)}_{k-1}] \cdot\cdot\cdot U[{(x)}_1,{(x)}_0]
         \Phi[z(0)]
    \nonumber  \\
 && ~~~~~\times ~ \frac{d{(x)}_0}{Q} \cdot\cdot\cdot \frac{d{(x)}_{k-1}}{Q}
\end{eqnarray}
where
$U^P[{(x)}_{j+1},{(x)}_j] \! = \! { \{ U^P[{(x)}_j,{(x)}_{j+1}] \} }^{-1}
 \! = \! U_{B'A'}^P[z(\tau)]$,
{}~$U[{(x)}_{j+1},{(x)}_j] = U_{B'A'}[z(\tau)]$ with $A'={(x)}_j$,
$B'={(x)}_{j+1}$, and $Q$ is a normalization constant. Set ${(x)}_k=y$,
$\tau_k = \tau$, ${(x)}_{k+1} = x$ and $\tau_{k+1} = \tau_k + \epsilon$.
Since $U^P[{(x)}_j,{(x)}_{j+1}]$ for each $j$, is a constant multiple of
the identity matrix, it follows from (25) that
\begin{equation}
  \psi'(x,\tau + \epsilon) = \frac{1}{Q}
                             \int U^P(y,x) U(x,y) \psi'(y,\tau) dy
\end{equation}

A curve $z(\tau)$ in $\cal M$ may be arbitrarily closely approximated by
$z_N(\tau)$ for large enough $N$, where $z_N(\tau_j)=z(\tau_j)$,
$j=0,1,...,N$; and in each of the intervals $[\tau_j,\tau_{j+1}]$,
$z_N(\tau)$ is the geodesic line. The element
$U^P(y,x) = U_{y,x}^P[z(\tau)]$ may be approximated by
\[
    U_{y,x}^P[z_N(\tau)] = \exp \left[ \; iS(x,y) \; \right]
\]
where $S(x,y)$ is Hamilton's principal function for a
`free' particle of mass $m$ from $x$ to a variable point $y$ (see also
Sec.2).
Here the
Lagrangian is $L^P$ with $\tau$ being the proper time of $\cal L$. The
action is given by [18]
\[
  S(x,y) = - \frac{m}{2\epsilon} g_{\mu\nu} \xi^\mu \xi^\nu
           - \frac{m}{2} \epsilon
\]
where $\xi^\mu = (x^\mu - y^\mu)$. Also, $U(x,y)$ is approximated by
$U_{x,y}[z_N(\tau)]$ up to the desired order which, from (9), is given by
\begin{eqnarray*}
   U_{x,y}[z_N(\tau)] &=& 1 + i \phi_\mu(x) \xi^\mu
    - \frac{1}{2} \left[ i\phi_{\mu,\nu} + \phi_\mu \phi_\nu \right]
      \xi^\mu \xi^\nu                            \\
   && ~~ + ~{\rm higher~order~terms}.
\end{eqnarray*}
Let $\psi(x,\tau) = \exp(im\tau/2)\psi'(x,\tau)$, then it follows from
(24) that
\begin{equation}
   \psi(x,0) = - \psi(x,2\pi/m),
\end{equation}
With the above substitutions, from (26), we have
\begin{equation}
  \psi(x,\tau+\varepsilon) = \frac{1}{Q} \int \exp
    \left[ -\frac{im}{2\epsilon} g_{\mu\nu} \xi^\mu \xi^\nu \right]
    U_{x,y}[z_N(\tau)] \psi(x-\xi,\tau) d\xi
\end{equation}
Eq.(28) holds exactly in the limit of infinite $N$, equivalently
$\epsilon=0$. As such it holds up to the first order in $\epsilon$,
which is sufficient for the present.

Expanding $\psi(x,\tau+\epsilon)$ and $\psi(x-\xi,\tau)$ in a Taylor
series about the point $(x,\tau)$ and comparing the coefficients of
$\epsilon^j$, $j=0,1$, yields $Q=-i(2\pi\epsilon/m)^2$ and
\begin{equation}
   i \frac{\partial \psi}{\partial \tau}
 = - \frac{1}{2m} \Pi_\mu \Pi^\mu \psi
\end{equation}
where $\Pi_\mu = (i \partial/\partial x^\mu \cdot 1 + \phi_\mu)$. In
view of the boundary condition (27), $\psi$ may be expressed as
\[
   \psi(x,\tau) = \sum_{-\infty}^{\infty} \psi_k(x) \omega_k(\tau)
\]
where for each $k$,
$\omega_k(\tau) = \sqrt{m/2\pi} \; \exp[i(k+1/2)m\tau]$ and the
$n$-vector $\psi_k$ satisfies
\begin{equation}
   \Pi_\mu \Pi^\mu \psi_k = (2k+1)m^2 \psi_k
\end{equation}
$k=0,\pm 1,\pm 2, \cdot\cdot\cdot$.
Eqs.(29) and (30) are remarkably similar to the corresponding
equations for the case
$n=1$ [9]. For $n=1$ and $k=0$, (30) reduces to the Klein-Gordon
equation in an electro-magnetic field.

Equation of motion (29) is an expected prescription to describe the
motion in a general gauge field on intuitive grounds that could be
inferred from the generalized Schr\"{o}dinger equation conjectured by
St\"{u}ckelberg [12,13]. The boundary condition (27) is a direct
result of the definition of the physical paths provided by (11).
As shown above, this boundary condition is crucial
in relating (29) to the Klein-Gordon equation. If all trajectories are
allowed to contribute, the resulting equation is still (29) but without
the boundary condition (27). Feynman [16] used this equation for $n=1$
to deduce the Klein-Gordon equation by restricting the solution to the
form $\psi_0(x) \omega_0(\tau)$.
Present treatment relates (29) with the Klein-Gordon equation (30) quite
naturally. Further to the arguments of Sec.3, this result provides
additional support for the assumption (11).
\begin{center}
{\bf 5. Concluding Remarks}
\end{center}

The variational principle determines a particle trajectory by requiring
the action to be stationary under all small deformations. In group
theoretical terms, this results in a requirement of equivalence between
the elements associated with a subset of the closed curves up to the
first order only. In this article, the classical action principle is
extended to require the equivalence of the global elements associated with
all of the curves. The resulting equation
selects an infinite subset, termed the physical paths, to assign to a
particle in motion.

Properties of the physical paths impart wave-like properties to a
particle in motion.
The wave-like behaviour of particles and the
multiplicity of allowed paths form the basis of the path integral
formulation. An imaginary value of $\alpha$ yielded by the present
extension, implies the compactness of the gauge groups which is
inherent in quantum mechanical equations in gauge fields.
Consequent description of the influence of the field enclosed by a closed
curve on the particles, as is the case with the Aharonov-Bohm effect, is
described by (11) to a large extent without an appeal to any other
theory. Thus the present formulation developes a coherent theory
unifying various treatments underlying the existing quantum mechanics.

The above results lead naturally to Feynman's path integral
formalism with physical paths being the contributing members.
The criterion imposed by (11)
on the physical paths plays a crucial role in the
deduction of the above results, some of which have been used to justify
the use of the path integral formalism. Thus the present formulation is
self-consistent.

In the present paper we have used a proper time-like parameter to
convert the problem of isolating the contribution from the physical
paths into a boundary condition on (29). This type of parameter was
introduced in a rather {\sl ad hoc} manner by several authors [12,18].
Here this parameter
gains a clearer physical significance. A need for a five-dimensional
relativistic wave equation has been felt for a long time, for the
existing equations suffer from some conceptual difficulties. In response
to this need, St\"{u}ckelberg originally conjectured the generalized
Schr\"{o}dinger equation for a particle in an Abelian gauge field [12,13].
There is a renewed interest in this equation to interpret it in a
more satisfactory framework than a conjecture, as well as to study its
implications (see e.g., [27]). Present formalism provides a
systematic derivation of the generalized Schr\"{o}dinger equation [9].
In the present article this derivation is generalized to
include a general gauge field.

In addition to accepting the conjecture of St\"{u}ckelberg [12,13],
Feynman selected a particular set of periodic solutions to
deduce the Klein-Gordon equation from the generalized Schr\"{o}dinger
equation. As pointed out above,
the physical paths are characterized by a boundary condition on (29).
This boundary condition confines the solution to a set described
by a class of periodic functions. As a consequence, the equation decomposes
into countably many four dimensional equations, one of them
being the Klein-Gordon equation. Thus the resulting boundary condition
provides an additional justification for the present treatment.

The techniques and results developed in the process of this work provide
also the ground work for a similar treatment of the spinors, and for an
extension of other theories based on the variational principle.
\newpage\noindent
\begin{center}
{\bf Acknowledgements}
\end{center}

The author is thankful to Professor H.O. Pritchard and Dr. C.C. Tai
for encouragement and discussions, to Dr. D. McConnell for
critical reading of the manuscript, and to Professor J. Marsden for
helpful comments.           \\[4ex]
\newpage\noindent
\begin{center}
{\bf References} \\[2.5ex]
\end{center}
\begin{enumerate}
\item H. Weyl, {\em Ann. Phys.} (Leipzig) {\bf 59}, 101 (1919).
\item H. Weyl, {\em Space-Time-Matter} (Translated by H.L. Brose).
Dover, New York. 1951. Ch.IV, Sec.35.
\item T. P. Chang and L. F. Li, {\em Am. J. Phys.} {\bf 56}, 586 (1988).
\item F. London, {\em Z. Phys.} {\bf 42}, 375 (1927).
\item C. N. Yang and R. L. Mills, {\em Phys. Rev.} {\bf 96}, 191 (1954).
\item C. N. Yang, {\em Phys. Rev. Lett.} {\bf 33}, 445 (1974).
\item L. D. Landau and E. M. Lifshitz,
{\em The Classical Theory of Fields} (Translated by M. Hamermesh).
Addison-Wesley, Reading, Massachusetts. 1965.  p.66.
\item S. K. Wong, {\em Nuovo Cimento} {\bf 65a}, 689 (1970).
\item S. R. Vatsya, {\em Can. J. Phys.} {\bf 67}, 634 (1989).
\item R. P. Feynman, {\em Rev. Mod. Phys.} {\bf 20}, 367 (1948).
\item R. P. Feynman and A. R. Hibbs,
{\em Quantum Mechanics and Path Integrals}.
McGraw-Hill, New York. 1965.
\item E. G. C. St\"{u}ckelberg, {\em Helv. Phys. Acta.} {\bf 14}, 322 (1941).
\item E. G. C. St\"{u}ckelberg, {\em Helv. Phys. Acta.} {\bf 15}, 23 (1942).
\item F. Guerra and R. Marra, {\em Phys. Rev. D.} {\bf 28}, 1916 (1983).
\item N. Cufaro-Petroni, C. Dewdney, P. R. Holland, A. Kyprianidis,
and  J. P. Vigier, {\em Phys. Rev. D.} {\bf 32}, 1375 (1985).
\item R. P. Feynman, {\em Phys. Rev.} {\bf 80}, 440 (1950).
\item H. Rund, {\em The Hamilton-Jocobi Theory in the Calculus of Variations}.
Van Nostrand, London. 1966.
\item V. Fock, {\em Phys. Zeit. Sowjet. un.} {\bf 12}, 404 (1937).
\item Y. Nambu, {\em Progr. Theo. Phys.} {\bf V}, 82 (1950).
\item N. E. Brali\'{c}, {\em Phys. Rev. D.} {\bf 22}, 3090 (1980).
\item H. Goldstein, {\em Classical Mechanics} (Second Edition).
Addison-Wesley, Reading, Massachusetts. 1980. pp. 365-367.
\item Y. Aharonov and D. Bohm, {\em Phys. Rev.} {\bf 115},485 (1959).
\item R. G. Chambers, {\em Phys. Rev. Lett.} {\bf 5}, 3 (1960).
\item A. Tonomura, N. Osakabe, T. Matsuda, T. Kawasaki and  J. Endo,
{\em Phys.Rev. Lett.} {\bf 56}, 792 (1986).
\item T. T. Wu and C. N. Yang, {\em Phys. Rev. D.} {\bf 12}, 3845 (1975).
\item P. A. M. Dirac, {\em The Principles of Quantum Mechanics}
(Fourth Edition). The Clarendon Press, Oxford, England. 1958. Sec.32.
\item A. Kyprianidis, {\em Phys. Rep.} {\bf 155}, 1 (1987).
\end{enumerate}
\end{document}